\documentclass[a4paper]{article}

\usepackage[dvips]{graphicx}

\usepackage{amsmath}
\usepackage{amsfonts}
\usepackage{amssymb}
\usepackage{amsthm}
\usepackage{amscd}
\usepackage[mathscr]{eucal}

\usepackage{url}
\usepackage{hyperref}

\newcommand{\myURL}[1]{
  \href{#1}{\url{#1}}
}

\newcommand{\myRef}[1]{
  \S\ref{#1}
}

\newlength{\facewd} \newlength{\faceht}

\theoremstyle{plain}

\theoremstyle{definition}

\newtheorem{definition}{Definition}[section]
\newtheorem{notation}{Notation}[section]
\newtheorem{remark}{Remark}[section]
\newtheorem{mrule}{Rule}[section]

\newenvironment{mySubequations}{
  \begin{subequations}
    
    }
  {
  \end{subequations}
  }

\newcommand{\droptext}[1]{\ensuremath{
    \text{\ #1\ }}
  }

\newcommand{\identity}[1]{\ensuremath{
    \langle #1 \rangle
    }
  }

\newcommand{\pubKey}[1]{\ensuremath{
    +{#1}
    }
  }

\newcommand{\privKey}[1]{\ensuremath{
    -{#1}
    }
  }

\newcommand{\encrypted}[2]{\ensuremath{
    \{{#1}\}_{#2}
    }
  }

\newcommand{\decrypted}[2]{\ensuremath{
    \{{#1}\}_{#2}^{-1}
    }
  }

\newcommand{\hashed}[1]{\ensuremath{
    H({#1})
    }
  }

\newcommand{\func}[1]{\ensuremath{
    F({#1})
    }
  }

\newcommand{\fresh}[1]{\ensuremath{
    \mathrel{\# \left( {#1} \right) }
    }
  }

\newcommand{\recognizable}[1]{\ensuremath{
    \mathrel{\phi \left( {#1} \right) }
    }
  }

\newcommand{\sees}{\ensuremath{
    \mathrel{\vartriangleleft}
    }
  }
\newcommand{\holds}{\ensuremath{
    \ni
    }
  }
\newcommand{\said}{\ensuremath{
    \mathrel{\mid\sim}
    }
  }
\newcommand{\believes}{\ensuremath{
    \mathrel{\mid\equiv}
    }
  }
\newcommand{\controls}{\ensuremath{
    \mathrel{\mid\Rightarrow}
    }
  }
\newcommand{\secret}[1]{\ensuremath{
    \mathrel{\stackrel{#1}{\leftrightarrow}}
    }
  }
\newcommand{\means}{\ensuremath{
    \rightsquigarrow
    }
  }

\newcommand{\project}[1]{\ensuremath{
    \mathop{\stackrel{#1}{\rightarrow}}
    }
  }

\newcommand{\eqdef}{\ensuremath{\triangleq}}

\begin{document}

\title{Transport Level Security: a proof using the Gong--Needham--Yahalom
  Logic}

\author{Walter D Eaves}

\date{\today}

\maketitle

\begin{abstract}
  This paper provides a proof of the proposed Internet standard Transport
  Level Security protocol using the Gong--Needham--Yahalom logic. It is
  intended as a teaching aid and hopes to show to students: the potency of
  a formal method for protocol design; some of the subtleties of
  authenticating parties on a network where all messages can be
  intercepted; the design of what should be a widely accepted standard.
\end{abstract}

\section{Transport Level Security Protocol}
\label{sec:auth}

This section provides an insight into the workings of the next
generation of authentication protocol: the Transport Level Security
Protocol version 1.0\cite{draft:tls}, the successor to the Secure
Sockets Layer\cite{draft:ssl}. To do this, the Gong--Needham--Yahalom,
GNY, logic \cite{GoNeYa90} is introduced which is a formal method for
proving the safety of a cryptographically-based protocol. It is
described at length in appendix \ref{cha:protocols}. When working
through protocols the relevant rule of inference will be stated and
will refer to those in the appendix.

The \emph{Transport Level Security handshake protocol}\cite{draft:tls},
TLS, has an unknown heritage, but it has a great deal of similarity to that
described in \cite{Denning:1981:TKD}. It is predicated on the existence of
readily available public keys: TLS's predecessor made use of X.509
certificates, see \cite{CCITTConsult88b}, issued by a Certification
Authority, CA, an example of which is \textit{Thawte}\cite{ca:thawte}. A
discussion of the limitations of certificate technology can be found in
R\"oscheisen's on--line paper \cite{Rosche95}.

TLS has three sub--protocols:

\begin{itemize}
\item Server anonymous
\item Server named, client anonymous
\item Server named, client named
\end{itemize}

\begin{figure}[htbp]
  \begin{center}
  \includegraphics{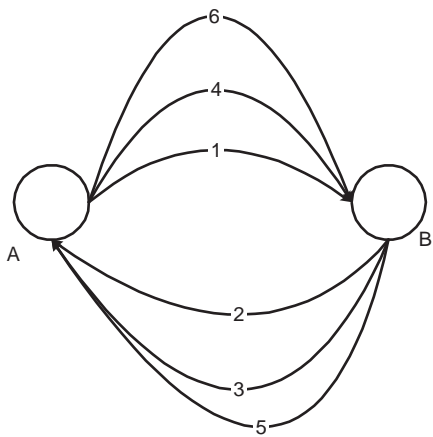}
  \caption{TLS protocol: Messages}
  \label{fig:auth-2}
  \end{center}
\end{figure}

These differ by who is required to send their X.509 certificates, the key
exchange protocol is different only when the client is named and thus has a
public--key that can be used. The messages are shown in figure
\ref{fig:auth-2} sent during a run of the protocol are more or less the
same for all sub--protocols. As can be seen, no key issuing server is
needed.

The TLS is a more complicated protocol than the \textit{Kerberos} which is
described in the appendix \myRef{sec:kerb}, before looking at TLS's
protocol proof, it might be best to examine \textit{Kerberos's}. Also,
there are some more examples of other authentication protocols being
investigated and found lacking\cite{Lowe96b}.

A protocol proof has three stages:
\begin{itemize}
\item Message Analysis
\item Pre--conditions Analysis
\item Belief deductions for each message
\end{itemize}

Message analysis involves formalizing the content of messages so that they
contain just keys and identifiers. Pre--conditions analysis formalizes what 
the parties to the protocol assume about the state of keys before
undertaking the protocol run. Belief deductions analyzes how each party can 
deduce new beliefs when it receives a new message.

\section{Messages for the Named Server Protocol}

There are six message exchanges. There is a provision for more, to settle
which cryptographic implementations to use and for the client to provide a
certificate, but this is the basic protocol for a named server to an
anonymous client.

\newcommand{\secretKey}{\ensuremath{
    \func{(N_A, T_A, N_B, T_B), {N'}_A}
    }
  }

\newcommand{\certificate}{\ensuremath{
    \encrypted{\pubKey{K_B}, \identity{B}, \identity{C}}{\privKey{K_C}}
    }
  }

\paragraph{Messages}

\begin{align}
  A \rightarrow B & \colon
  (N_A, T_A)
  \label{eq:tls:m1} \tag{$M_1$} \\
  B \rightarrow A & \colon
  (N_B, T_B)
  \label{eq:tls:m2} \tag{$M_2$} \\
  B \rightarrow A & \colon
  \certificate
  \label{eq:tls:m3} \tag{$M_3$} \\
  A \rightarrow B & \colon
  \encrypted{{N'}_A}{\pubKey{K_B}}
  \label{eq:tls:m4} \tag{$M_4$} \\
  B \rightarrow A & \colon 
  \encrypted{H(K_{AB}, AB_5, (M_1, M_2, M_3, M_4))}{K_{AB}}
  \label{eq:tls:m5} \tag{$M_5$} \\
  A \rightarrow B & \colon 
  \encrypted{H(K_{AB}, AB_6, (M_1, M_2, M_3, M_4))}{K_{AB}}
  \label{eq:tls:m6} \tag{$M_6$} \\
  \intertext{where}
  K_{AB} & = \secretKey \notag \\
  \intertext{and, the behaviour of the public and private keys is:}
  \encrypted{\encrypted{X}{\privKey{K}}}{\pubKey{K}} & = X \notag \\
  \encrypted{\encrypted{X}{\pubKey{K}}}{\privKey{K}} & = X \notag
\end{align}

and
\begin{equation*}
  \begin{aligned}
    AB_5 \eqdef \droptext{``server finished''}
  \end{aligned}
  \droptext{,}
  \begin{aligned}
    AB_6 \eqdef \droptext{``client finished''}
  \end{aligned}
\end{equation*}

The messages can be summarized as follows:

\begin{description}
\item[\ref{eq:tls:m1}] $A$ sends a timestamp and a nonce to $B$.
\item[\ref{eq:tls:m2}] $B$ sends another timestamp and another nonce to $A$.
\item[\ref{eq:tls:m3}] $B$ sends its certificate signed by the
  certification authority; it contains $\pubKey{K_B}$, $B$'s public--key.
\item[\ref{eq:tls:m4}] $A$ returns the ``pre--master secret'' ${N'}_A$
  encrypted under $\pubKey{K_B}$.
\item[\ref{eq:tls:m5}] $B$ sends a hash of the session key, a tag
  indicating the protocol stage $AB_5$\footnote{Actually all stages of the
    protocol are marked with a stage identifier, but it is not necessary to
    consider all of them.} and all preceding messages exchanged to $A$.
\item[\ref{eq:tls:m6}] $A$ sends a hash of the session key, a tag
  indicating the protocol stage $AB_6$ key and all preceding messages to
  $B$.
\end{description}

\section{Pre--Conditions}

\begin{mySubequations}

\paragraph{Certificates}

\begin{enumerate}
  
\item Some Expectations

  $A$ expects to use $C$ as the certification authority and expects to use
  $B$, so
  \begin{equation}
    \label{eq:tls:c:0}
    \begin{aligned}
      A & \holds \identity{C}
    \end{aligned}
    \quad
    \begin{aligned}
      A & \holds \identity{B}
    \end{aligned}
  \end{equation}

\item Using Them
  
  The role of the unseen certification authority, $C$, is pivotal. Even
  though $C$ has used the private key $\privKey{K_C}$ to create the
  certificate, this key is not used again.

  \begin{equation}
    \begin{aligned}
      B \believes A \believes \project{\pubKey{K_C}} C
    \end{aligned}
    \text{,} \quad
    \begin{aligned}
      A \believes \project{\pubKey{K_C}} C
    \end{aligned}
    \label{eq:tls:c:1}
  \end{equation}
  
  The two parties rely on the public--key being available. $A$ must be able
  to get $K_C$:

  \begin{equation}
    \begin{aligned}
      A \holds \pubKey{K_C}
    \end{aligned}
    \text{,} \quad
    \begin{aligned}
      B \believes A \holds \pubKey{K_C}
    \end{aligned}
    \label{eq:tls:c:2}
  \end{equation}
  
\item Trusting in them
  
  The pre--conditions regarding $B$'s certificate are as follows. $A$ and
  $B$ both trust $C$ to deliver the correct identity with the public key.

  \begin{equation}
    \begin{aligned}
      A \believes C \controls ( \pubKey{K_P}, \identity{P} )
    \end{aligned}
    \text{,} \quad
    \begin{aligned}
      B \believes C \controls ( \pubKey{K_P}, \identity{P} )
    \end{aligned}
    \label{eq:tls:c:3}
  \end{equation}
  
\item Meaning of the contents
  
  For $A$ the assumption underlying a certificate is that the public--key is 
  the public--key of the named party.
  \begin{equation}
    A \believes ( \pubKey{K_P}, \identity{P} ) \means
    C \believes \project{\pubKey{K_P}} P
    \label{eq:tls:c:4}
  \end{equation}

  And $A$ believes $C$ when $C$ names a key:
  \begin{equation}
    \label{eq:tls:c:5}
    A \believes C \controls \project{\pubKey{K_P}} P
  \end{equation}

\end{enumerate}

\paragraph{System Capabilities}

\begin{enumerate}

\item $B$ believes that $A$ can generate a nonce and keep it secret to pass 
  it on as the pre--master secret.

  \begin{equation}
    \begin{aligned}
      B \believes A \controls \fresh{X}
    \end{aligned}
    \text{,} \quad
    \begin{aligned}
      B \believes A \secret{X} A
    \end{aligned}
    \label{eq:tls:c:6}
  \end{equation}
  
\item $A$ and $B$ have both assumed that the other can generate the master
  secret if presented with the components.

  \begin{equation}
    \begin{aligned}
      A \believes B \controls \func{X,Y}
    \end{aligned}
    \text{,} \quad
    \begin{aligned}
      B \believes A \controls \func{X,Y}
    \end{aligned}
    \label{eq:tls:c:7}
  \end{equation}

  and, of course, they do

  \begin{equation}
    \begin{aligned}
      B \controls \func{X,Y}
    \end{aligned}
    \text{,} \quad
    \begin{aligned}
      A \controls \func{X,Y}
    \end{aligned}
    \label{eq:tls:c:8}
  \end{equation}

\item $B$ has a private--key and holds his own certificate.
  
  \begin{equation}
    \begin{aligned}
      B \holds \privKey{K_B}
    \end{aligned}
    \text{,} \quad
    \begin{aligned}
      B \holds \certificate
    \end{aligned}
    \label{eq:tls:c:9}
  \end{equation}

  Note that the format of a certificate does includes a statement of the
  identity of $C$. Although not used in this protocol it is an important
  part of it since it allows the public--key of $C$ to be checked.

\end{enumerate}

\end{mySubequations}

\section{Belief Deductions Analysis}

\begin{mySubequations}

\begin{enumerate}
    
\item Messages \ref{eq:tls:m1} received by $B$ and \ref{eq:tls:m2} received 
  by $A$
  
  These nonce and timestamp exchanges are important, because they are used
  in the generation of the key. The vindication is the appearance of the
  time--stamp, which is definitely fresh, and the the rule
  \eqref{eq:gny:f1} freshens the nonces.

\begin{equation}
  \begin{aligned}
    A & \holds M_1, M_2 \droptext{By \eqref{eq:gny:p1}} \\
    A & \holds N_B, T_B, N_A, T_A \\
    A & \believes \fresh{N_B}
  \end{aligned}
  \droptext{and} 
  \begin{aligned}
    B & \holds M_1, M_2 \droptext{By \eqref{eq:gny:p1}} \\
    B & \holds N_A, T_A, N_B, T_B \\
    B & \believes \fresh{N_A}
  \end{aligned}
  \label{eq:tls:a:1}
\end{equation}

\item Message \ref{eq:tls:m3} received by $A$
  
  $A$ receives the certificate from, presumably, $B$. By
  \eqref{eq:tls:c:1}, \eqref{eq:tls:c:2} and \eqref{eq:gny:i4}, $A$ now
  has:

  \begin{equation*}
    A \believes C \said \certificate
  \end{equation*}

  and can decrypt the contents to discover:

  \begin{equation*}
    C \said  ( \pubKey{K_B}, \identity{B} )
  \end{equation*}

  By \eqref{eq:tls:c:3} and \eqref{eq:tls:c:4}

  \begin{equation*}
    A \believes C \believes \project{\pubKey{K_B}} B
  \end{equation*}

  By \eqref{eq:tls:c:5} and \eqref{eq:gny:j1}

  \begin{equation*}
    A \believes \project{\pubKey{K_B}} B
  \end{equation*}

  And of course
  \begin{equation}
    \begin{split}
      A \holds M_3 & \droptext{By \eqref{eq:gny:p1}.} \\
      A \holds M_4 & \droptext{Because $A$ creates it.}
    \end{split}
    \label{eq:tls:a:3}
  \end{equation}

  Notice that $A$ does not know that the sender of this message was $B$.
  
\item Message \ref{eq:tls:m4} received by $B$
  
  $B$ gains knowledge of the following:
  \begin{equation}
    \begin{split}
      B \holds M_3 & \droptext{Since it sent it} \\
      B \holds M_4 & \droptext{By \eqref{eq:gny:p1}}
    \end{split}
    \label{eq:tls:a4}
  \end{equation}

  From which by \eqref{eq:gny:i2} where $\identity{S}$ is ${N'}_A$.
  \begin{align*}
    B & \sees {N'}_A \\
    B & \holds {N'}_A
  \end{align*}
  This is something of an innovation: ${N'}_A$ has not been established as
  a shared secret by the conventional method of passing it along a channel
  secured by a long--term key or comparing it to a pre--stored hash, but it
  \emph{will} be established as a secret by the subsequent correct
  operation of the protocol.
  
  Since ${N'}_A$ came under cover of the public key and it will be later
  identified as unique to the sender, it is a shared secret, so by
  \eqref{eq:tls:c:6}.
  \begin{align*}
    B & \believes A \secret{{N'}_A} B  \\
    B & \believes \fresh{{N'}_A}
  \end{align*}

  Now by \eqref{eq:tls:c:8} and \eqref{eq:tls:a:1}
  \begin{align*}
    B & \holds K_{AB} \\
    B & \believes \fresh{K_{AB}} \\
    \because B & \holds N_A, N_B, T_A, T_B
  \end{align*}

  $B$ can now construct the response message:
  \begin{align*}
    B & \holds M_1, M_2, M_3, M_4  \\
    B & \holds H(K_{AB}, AB_5, (M_1, M_2, M_3, M_4))
  \end{align*}
  
\item Message \ref{eq:tls:m5} received by $A$
  
  $A$ now receives a message from $B$ which can only be understood
  \emph{correctly}, if both $A$ and $B$ have agreed upon $K_{AB}$.

  $A$ performs some pre--calculations:
  \begin{align*}
    A & \holds K_{AB} \\
    \because A & \holds {N'}_A, N_A, N_B, T_A, T_B \\
    A & \holds H(K_{AB}, AB_5, (M_1, M_2, M_3, M_4))
  \end{align*}
  Because $A$ has been collecting the messages as well \eqref{eq:tls:c:8}
  and holding all previous messages \eqref{eq:tls:a:1} and
  \eqref{eq:tls:a:3}.

  By \eqref{eq:gny:i1} $A$ can decrypt the message
  \begin{align*}
    A \sees & H(K_{AB}, AB_5, (M_1, M_2, M_3, M_4)) \\
    \therefore & \\
    A \believes & H(K_{AB}, AB_5, (M_1, M_2, M_3, M_4)) \\
    & \means B \holds H(K_{AB}, AB_5, (M_1, M_2, M_3, M_4)) 
  \end{align*}

  $A$ can now make a series of justified conclusions, by \eqref{eq:gny:i3}
  \begin{align*}
    A & \believes B \holds {N'}_B, K_{AB} \\
    A & \believes B \holds M_1, M_2, M_3, M_4 \\
    A & \believes B \holds N_A, T_A, N_B, T_B 
  \end{align*}

  $A$ can now validate the identity of the other party:
  \begin{equation}
    \begin{aligned}
      A & \believes B \said (N_B, T_B)  \\
      A & \believes B \believes A \said (N_A, T_A)  \\
      A & \believes B \believes A \secret{K_{AB}} B \\
      A & \believes B \believes \fresh{K_{AB}}
    \end{aligned}
    \text{\quad}
    \begin{aligned}
      A & \believes B \sees M_1 \\
      A & \believes B \said M_2 \\
      A & \believes B \said M_3 \\
      A & \believes B \sees M_4
    \end{aligned}
    \label{eq:tls:a:5}
  \end{equation}
  
  It should be clear now why the key $K_{AB}$ is hashed into the hash
  signature $H(K_{AB}, AB_5, (M_1, M_2, M_3, M_4))$. A hash is only
  validated by inclusion of a secret and a nonce, see \eqref{eq:gny:i3},
  the key $K_{AB}$ is both.

\item Message \ref{eq:tls:m6} received by $B$
  
  By a similar argument to that used for $A$, it is clear:

  \begin{equation}
    \begin{aligned}
      B & \believes A \said {N'}_A \\
      B & \believes A \believes B \said (N_B, T_B) \\
      B & \believes A \believes A \secret{K_{AB}} B \\
      B & \believes A \believes \fresh{K_{AB}}
    \end{aligned}
    \text{\quad}
    \begin{aligned}
      B & \believes A \said M_1 \\
      B & \believes A \sees M_2 \\
      B & \believes A \sees M_3 \\
      B & \believes A \said M_4
    \end{aligned}
  \end{equation}
  
  Since $A$ could only generate this message if in possession of $K_{AB}$,
  $B$ can deduce that $A$ is the party with whom it shares the key and the
  whole protocol run is current.

\end{enumerate}

\end{mySubequations}

\section{Summary: Innovations and Possible Attacks}

\paragraph{Summary}

The Transport Level Security handshake protocol is quite ingenious: it lets
$A$ send a random message under a public key which is used as an
identifying secret shared by the parties before it has been established as
such. A challenge and response protocol, the challenge is issued in
plain--text and the response returns it as cipher--text, so that the
challenger can verify that the responder knows the shared session key. This
protocol is effectively a challenge and response protocol with the
generated session key, which is created from the secret sent under the
public--key.

The protocol is also exemplary in its use of stage identifiers and hash
digests. The stage identifiers change the hash digest between messages
\ref{eq:tls:m5} and \ref{eq:tls:m6}. The hash digests validate the
whole protocol run.

\paragraph{Attacks}

The critical point of the protocol is the transmission of the public--key
with which the client should respond with a nonce encrypted under it. The
man-in-the-middle attack is well known here: all that need be done is to
intercept \ref{eq:tls:m3} and substitute a bogus certificate. The fraud
then hinges upons the expectations of $A$, \eqref{eq:tls:c:0}. 

\begin{enumerate}

\item Impersonate $C$ and $B$
  
  If $A$ is not expecting to use $C$ or $B$ then the attacker, $M$, can
  substitute a different certificate for another service: $D$.
  
\item Impersonate $B$

  If $A$ is not expecting to use $B$ but is expecting
  a certificate issued by $C$ then $M$ can create a service $D$ and attempt
  to have $C$ issue a certificate for it.

\end{enumerate}

It is quite easy to provide a service that looks like $B$ and appears to be
at the address of $B$ this is rather more difficult with a certification
authority because the public certification scheme proposed in
\cite{OSI:dir} is based upon the following:

\begin{itemize}
\item Certification authorities are well--known in that their addresses and 
  public--keys can be obtained from many sources.
\item Certificates contain lists of certification authorities which allow a 
  client to match known certification authorities to those found in the
  certificates.
\end{itemize}

Certification authorities currently do nothing other than provide
certificates, so all a client obtains from a certificate is some
accountability. If defrauded the client can attempt to locate the server
who perpetrated the fraud.

\paragraph{Another Useful Feature}

One of the provisions of TLS is to allow the server to pass to the client
another key to use in place of the certificate key. This may be necessary
for any of the following reasons:

\begin{itemize}
\item The client lacks an implementation to encrypt with the server's
  public--key.
\item The server does not wish to use its public--key.
\item Restrictions on key size require that a smaller or larger key must be
  used.
\end{itemize}

The client would receive a different public--key but that must be signed
under the certificate key for the client to have any faith in it. If the
client had chosen to use the alternative key because it lacked an
implementation, it would still need to decrypt under the certification
authority's key, it is unlikely that the client would be able to do this
and not make use of the server's certificate key.

The alternative cryptosystem to system used for certificates is the
Diffie--Hellman public--key system\cite{crypto:dh}.

\section{Other sub-protocols}

The protocol described above was the named server protocol, where the
server must provide a certificate. There are two other sub--protocols.

\subsection{Anonymous Server}

In this variant, the server is anonymous and creates a public and private
key pair to be used to establish the session key. It would usually use the
Diffie--Hellman scheme in this case and would simply send to the client the
public key instead of the certificate. This does not weaken the protocol at
all, the client and the server will be able to mutually authenticate one
another, but the server is unknown to the client and to a certification
authority. There is no chain of accountability that could help to locate a
fraudulent server.

\subsection{Named Client and Server}

This variant provides some accountability to the server of the client's
identity and it relies upon the client having a certificate. The protocol
is the same as the named server protocol, but the server can request a
certificate from the client prior to the client sending the pre--master
secret. If the client has no certificate it replies by returning no
certificates, whereupon the server can take its own action, which may well
be to raise an error and not complete the protocol.

\section{Summary}

TLS, like its predecessor the Secure Socket Layer, SSL, does provide both
parties with a mutual belief that the shared session key is a fresh secret.
It also, like SSL, can provide the client with some account of the server's
Internet location and, unlike its predecessor, it does support mutual
authentication certificate exchange. Suffice to say that identities can be
securely established---using X.509 certificates---and that a session key
can be securely established.

A protocol proof is just a basis for a secure implementation. The software
engineering of the authentication protocol has to be considered. An example
of such a failure to ensure that a software implementation was invulnerable
to attack can be found at \cite{sec:pkcs}. The problem with that
implementation was that error messages proved to be too informative
allowing a sophisticated intruder to recover a session key more quickly
than by key trial. Lowe's paper \cite{Lowe96b} has some other
implementation attacks.


\newpage
\appendix

\section{Appendix: Gong--Needham--Yong Logic}
\label{cha:protocols}

This is only a cursory introduction to this simple proof system, despite
stimulating a great deal of research interest it is relatively unchanged.

\subsection{Brief History}

Authentication protocols had been developed and discussed, in particular
the CCITT X.509 protocol of 1987 \cite{CCITTConsult88b} was the source for
some debate and it was \cite{BurrowsAbadiNeedham90}, which proved a
weakness existed in it. The protocol was extended by \cite{GoNeYa90} and it 
has been adapted and used in other contexts, \cite{sec:dist}. It may even
have been superceded by a calculus that is somewhat less intuitive
\cite{sec:spi}. It is now used principally to illustrate that there is more
to secure information that just have believing it to be so, for which see
\cite{sec:xu} and \cite{Lowe96b}.

\subsection{The Logic}

\begin{notation}[Formulae] Formulae is the name used to refer to a
  bit--string. Certain useful operations can be applied to bit strings,
  which are given below.  $X$ and $Y$ range over formulae; $S$ over
  formulae that are secrets and $K$ over formulae that are keys.
  \begin{description}
  \item[$(X, Y)$] Resulting bit--string is a concatenation of two formalue.
  \item[$\encrypted{X}{K}$ and $\decrypted{X}{K}$] Results are bit--strings
    are $X$ encrypted and decrypted under a symmetric cryptosystem
    with key $K$, respectively.
  \item[$\encrypted{X}{\pubKey{K}}$ and $\encrypted{X}{\privKey{K}}$]
    Bit--string results are $X$ encrypted under the public key and under
    the private key, respectively, of an asymmetric cryptosystem with
    public key $\pubKey{K}$ and and private key $\privKey{K}$.
  \item[$\hashed{X}$] Result is $X$ after having been subjected to a
    one-way function.
  \item[$\func{X_1, X_2, \cdots, X_n}$] Bit--string is the result after
    applying the many-to-one function $F$, which is an invertible and
    computationally feasible in both directions, to all of $X_1, X_2,
    \cdots, X_n$.
  \end{description}
\end{notation}

\begin{notation}[Statements] Statements make an assertion about a
  property of a formula. $P$ and $Q$ denote principals---clients, agents
  and servers: $X$ is a message; $K$ a key; $S$ a secret.
  \begin{description}
  \item[$P \sees X$] $P$ is told the message $X$.
  \item[$P \holds X$] $P$ holds or can obtain the message $X$.
  \item[$P \said X$] $P$ has once conveyed $X$.
  \item[$P \believes X$] $P$ believes the message $X$.
  \item[$P \believes \fresh{X}$] $P$ believes that $X$ is a fresh
    statement, not seen before in this run of the protocol. $X$ is often
    known as a nonce. (Note freshness is a belief relative to a
    principal). 
  \item[$P \believes \recognizable{X}$] $P$ believes that $X$ is
    recognizable: $P$ is able to decode $X$ it has a recognizable transfer
    syntax. (Same note as above).
  \item[$P \believes P \secret{S} Q$] $P$ believes it shares the secret $S$ 
    with $Q$.
  \item[$P \believes \project{\pubKey{K}} Q$] $P$ believes that
    $\pubKey{K}$ is the public key of $Q$.
  \item[$P \believes Q \controls C$] $P$ believes that $Q$ has jurisdiction 
    over $C$.
  \item[$P \believes Q \controls Q \believes *$] $P$ believes that $Q$ has
    jurisdiction over all of beliefs held by $P$.
  \item[$X \means C$] $X$ is a message expressing the statement $C$.
  \item[$C_1, C_2$] Conjunction of two statements: $C_1$ and $C_2$.
  \item[$*X$] The message did not originate from its current location.
  \end{description}
\end{notation}

\begin{remark}[Epochs]
  Time is divided into two epoch: past and present. The present is the
  run of a protocol. The past is all other runs of protocols. $P
  \believes X$, if a pre--condition, is valid for all of the present.
  Beliefs held in the past are not necessarily carried forward to the
  present.
\end{remark}

\begin{remark}[Encryption] There are some assumptions about encrypted
  messages: 

  \begin{enumerate}
    \item Messages are assumed to be encrypted as a whole.
    \item For recipients: each encrypted message contains enough
      redundancy to allow the recipient to determine, on decryption,
      that the right key has been used to do so.
    \item For senders: each message contains enough information to
      allow senders to detect and ignore messages that originated
      from them.
  \end{enumerate}
  Also,
  \begin{enumerate}
  \item The key cannot be deduced from the encrypted message.
  \item The message can be understood by only those who possess the
    correct decrypting key.
  \end{enumerate}
\end{remark}

The logic of authentication is a set of inference rules. The premisses 
are stated above the deductive line, the conclusion below.

\begin{mrule}[Universal--Local] This is an axiom from modal logic.
  \begin{equation}
    \label{eq:gny:localize}
    \tag{Localize}
    \frac{C_1}{C_2} \Rightarrow \frac{P \believes C_1}{P \believes C_2}
  \end{equation}
\end{mrule}

\begin{mrule}[Being--Told] The rules about the ``being-told'' operator
  $\sees$. 

  \begin{equation}
    \frac{P \sees *X}{P \sees X}
    \label{eq:gny:t1}
    \tag{T1} 
  \end{equation}
  \eqref{eq:gny:t1} says that if one is in receipt of a message
  that did originate from elsewhere, one is still aware of it.

  \begin{equation}
    \frac{P \sees (X,Y)}{P \sees X}
    \label{eq:gny:t2}
    \tag{T2} 
  \end{equation}
  \eqref{eq:gny:t2} says that if one is in receipt of a composite message
  one is in receipt of each part of it.

  \begin{equation}
    \frac{P \sees \encrypted{X}{K}, P \holds K}{P \sees X}
    \label{eq:gny:t3}
    \tag{T3} 
  \end{equation}
  \eqref{eq:gny:t3} is simple one must possess the right key to understand
  encrypted messages.

  \begin{equation}
    \frac{P \sees \encrypted{X}{\pubKey{K}},
      P \holds \privKey{K}}
    {P \sees X}
    \label{eq:gny:t4}
    \tag{T4} 
  \end{equation}
  \eqref{eq:gny:t4} is the same statement for public--key systems,
  decrypting with the private key.

  \begin{equation}
    \frac{P \sees \func{X,Y}, P \holds X}{P \sees Y}
    \label{eq:gny:t5}
    \tag{T5} 
  \end{equation}
  \eqref{eq:gny:t5} is the same statement for a combination function.

  \begin{equation}
    \frac{
      \encrypted{
        \encrypted{X}{\privKey{K}}
          }{\pubKey{K}} = X, 
        P \sees \encrypted{X}{\privKey{K}}, 
        P \holds \pubKey{K}
        } {P \sees X}
    \label{eq:gny:t6}
    \tag{T6}
  \end{equation}
  \eqref{eq:gny:t6} is the same statement for public--key systems, but
  decrypting with the public key. Note with this rule, the requirement that
  encryption with a public--key and then with the private--key yields the
  original message. Not all asymmetric cryptosystems have this property.
\end{mrule}

\begin{mrule}[Possession] Rules for the $\holds$ operator:

  \begin{equation}
    \frac{P \sees X}{P \holds X}
    \label{eq:gny:p1}
    \tag{P1} 
  \end{equation}
  \eqref{eq:gny:p1} states that one can possess what one is told.

  \begin{equation}
    \frac{P \holds X, P \holds Y}{P \holds (X, Y), P \holds F(X, Y)}
    \label{eq:gny:p2}
    \tag{P2} 
  \end{equation}
  \eqref{eq:gny:p2} states that if in possession of two messages one can
  create a concatenation and apply a function to them. 

  \begin{equation}
    \frac{P \holds (X, Y)}{P \holds X} 
    \label{eq:gny:p3}
    \tag{P3} 
  \end{equation}
  \eqref{eq:gny:p3} possession of a composite yields possession of the
  components. 

  \begin{equation}
    \frac{P \holds X}{P \holds \hashed{X}}
    \label{eq:gny:p4}
    \tag{P4} 
  \end{equation}
  \eqref{eq:gny:p4} possession of a message allows the hash of
  it to be generated.

  \begin{equation}
    \frac{P \holds F(X,Y), P \holds X}{P \holds Y}
    \label{eq:gny:p5}
    \tag{P5} 
  \end{equation}
  \eqref{eq:gny:p5} for the combination function $F()$, given $X$, $Y$ can
  be determined.

  \begin{equation}
    \frac{P \holds K, P \holds X}
    {P \holds \encrypted{X}{K}, P \holds \decrypted{X}{K}}
    \label{eq:gny:p6}
    \tag{P6} 
  \end{equation}
  \eqref{eq:gny:p6} one can encrypte and decrypt with key
  and message.

  \begin{equation}
    \frac{P \holds \pubKey{K}, P \holds X}
    {P \holds \encrypted{X}{\pubKey{K}}}
    \label{eq:gny:p7}
    \tag{P7} 
  \end{equation}
  \eqref{eq:gny:p7} encryption under public--key.

  \begin{equation}
    \frac{P \holds \privKey{K}, P \holds X}
    {P \holds \encrypted{X}{\privKey{K}}}
    \label{eq:gny:p8}
    \tag{P8} 
  \end{equation}
  \eqref{eq:gny:p8} encryption under private--key.
\end{mrule}

\begin{mrule}[Freshness] These rules specify what can be deduced from fresh 
  messages.
  \begin{equation}
    \frac{P \believes \fresh{X}}{
      P \believes \fresh{X, Y},
      P \believes \fresh{\func{X}}
      }
    \label{eq:gny:f1} \tag{F1} 
  \end{equation}

  \begin{equation}
    \frac{P \believes \fresh{X}, P \holds K}{
      P \believes \fresh{\encrypted{X}{K}},
      P \believes \fresh{\decrypted{X}{K}}
      }
    \label{eq:gny:f2} \tag{F2} 
  \end{equation}

  \begin{equation}
    \frac{P \believes \fresh{X}, P \holds \pubKey{K}}{
      P \believes \fresh{\encrypted{X}{\pubKey{K}}}
      }
    \label{eq:gny:f3} \tag{F3} 
  \end{equation}

  \begin{equation}
    \frac{P \believes \fresh{X}, P \holds \privKey{K}}{
      P \believes \fresh{\encrypted{X}{\privKey{K}}}
      }
    \label{eq:gny:f4} \tag{F4} 
  \end{equation}

  \begin{equation}
    \frac{P \believes \fresh{\pubKey{K}}}{
      P \believes \fresh{\privKey{K}}
      }
    \label{eq:gny:f5} \tag{F5} 
  \end{equation}

  \begin{equation}
    \frac{P \believes \fresh{\privKey{K}}}{
      P \believes \fresh{\pubKey{K}}
      }
    \label{eq:gny:f6} \tag{F6} 
  \end{equation}

  \begin{equation}
    \frac{P \believes \recognizable{X},
      P \believes \fresh{K}, P \holds K }{
      P \believes \fresh{\encrypted{X}{K}},
      P \believes \fresh{\decrypted{X}{K}}
      }
    \label{eq:gny:f7} \tag{F7} 
  \end{equation}

  \begin{equation}
    \frac{P \believes \recognizable{X},
      P \believes \fresh{\pubKey{K}}, P \holds \pubKey{K} }{
      P \believes \fresh{\encrypted{X}{\pubKey{K}}}
      }
    \label{eq:gny:f8} \tag{F8} 
  \end{equation}

  \begin{equation}
    \frac{P \believes \recognizable{X},
      P \believes \fresh{\privKey{K}}, P \holds \privKey{K} }{
      P \believes \fresh{\encrypted{X}{\privKey{K}}}
      }
    \label{eq:gny:f9} \tag{F9} 
  \end{equation}

  \begin{equation}
    \frac{P \believes \fresh{X}, P \holds X}{P \believes \fresh{\hashed{X}}} 
    \label{eq:gny:f10} \tag{F10} 
  \end{equation}

  \begin{equation}
    \frac{P \believes \fresh{\hashed{X}},
      P \holds \hashed{X}}{P \believes \fresh{X}} 
    \label{eq:gny:f11} \tag{F11} 
  \end{equation}
\end{mrule}

\begin{mrule}[Recognizability] When one can claim a formula is
  recognizable.
  \begin{equation}
    \frac{P \believes \recognizable{X}}{
      P \believes \recognizable{X, Y}, P \believes{\func{X}}
      }
    \label{eq:gny:r1} \tag{R1} 
  \end{equation}

  \begin{equation}
    \frac{P \believes \recognizable{X}, P \holds K}{
      P \believes \recognizable{\encrypted{X}{K}},
      P \believes \recognizable{\decrypted{X}{K}}
      }
    \label{eq:gny:r2} \tag{R2} 
  \end{equation}

  \begin{equation}
    \frac{P \believes \recognizable{X}, P \holds \pubKey{K}}{
      P \believes \recognizable{\encrypted{X}{\pubKey{K}}}
      }
    \label{eq:gny:r3} \tag{R3} 
  \end{equation}

  \begin{equation}
    \frac{P \believes \recognizable{X}, P \holds \privKey{K}}{
      P \believes \recognizable{\encrypted{X}{\privKey{K}}}
      }
    \label{eq:gny:r4} \tag{R4} 
  \end{equation}

  \begin{equation}
    \frac{P \believes \recognizable{X}, P \holds X}{
      P \believes \recognizable{\hashed{X}}
      }
    \label{eq:gny:r5} \tag{R5} 
  \end{equation}

  \begin{equation}
    \frac{P \holds \hashed{K}}{
      P \believes \recognizable{X}
      }
    \label{eq:gny:r6} \tag{R6}
  \end{equation}
\end{mrule}

\begin{mrule}[Message Interpretation] A secret $S$ used for identification
  is denoted $\identity{S}$---this is to allow it to be distinguished from other
  secrets that might be in the message.
  \begin{equation}
    \frac{P \sees *\encrypted{X}{K}, P \holds K,
      P \believes P \secret{K} Q,
      P \believes \recognizable{X}, P \believes \fresh{X, K}}{
      P \believes Q \said X, P \believes Q \said \encrypted{X}{K}, P
      \believes Q \holds K
      }
    \label{eq:gny:i1} \tag{I1} 
  \end{equation}
  This specifies the flow of beliefs on receipt of an encrypted message
  under a shared--key cryptosystem. Notice that $P \believes \fresh{X, K}$, 
  either the key is fresh or the message is fresh. Usually the message is
  freshened by adding a nonce (or timestamp).

  \begin{equation}
    \frac{
      \begin{matrix}
        P \sees & *\encrypted{X, \identity{S}}{\pubKey{K}}, \\
        P \holds & (\privKey{K}, S), \\
        P \believes & \project{\pubKey{K}} P, \\
        P \believes & P \secret{S} Q, \\
        P \believes & \recognizable{X, S}, \\
        P \believes & \fresh{X, S, \pubKey{K} }
      \end{matrix}
      } {
      \begin{matrix}
        P \believes & Q \said (X, \identity{S}), \\
        P \believes & Q \said \encrypted{X, \identity{S}}{\pubKey{K}}, \\
        P \believes & Q \holds \pubKey{K} 
      \end{matrix}
      }
    \label{eq:gny:i2} \tag{I2} 
  \end{equation}
  Message sent encrypted under a public--key. Normally such messages are
  anonymous, since anyone can use the public--key, but an identifying
  secret $S$ is passed.

  \begin{equation}
    \frac{P \sees *\hashed{X, \identity{S}},
      P \holds (X, S),
      P \believes P \secret{S} Q, 
      P \believes \fresh{X, S} } {
      P \believes Q \said (X, \identity{S}),
      P \believes Q \said \hashed{X, \identity{S}}
      }
    \label{eq:gny:i3} \tag{I3} 
  \end{equation}
  Passing a hashed message.

  \begin{equation}
    \frac{P \sees \encrypted{X}{\privKey{K}},
      P \holds \pubKey{K},
      P \believes \project{\pubKey{K}} Q, 
      P \believes \recognizable{X} } {
      P \believes Q \said X,
      P \believes Q \said \encrypted{X}{\privKey{K}}
      }
    \label{eq:gny:i4} \tag{I4} 
  \end{equation}
  Passing a message encrypted under a private--key does \emph{not} require
  an identifying secret.

  \begin{equation}
    \frac{P \sees \encrypted{X}{\privKey{K}},
      P \holds \pubKey{K}, P \believes \project{\pubKey{K}} Q,
      P \believes \recognizable{X},
      P \believes \fresh{X, \pubKey{K}} } {
      P \believes Q \holds (\privKey{K}, X)
      }
    \label{eq:gny:i5} \tag{I5} 
  \end{equation}

  \begin{equation}
    \frac{P \believes Q \said X, P \believes \fresh{X}}{
      P \believes Q \holds X
      }
    \label{eq:gny:i6} \tag{I6} 
  \end{equation}

  \begin{equation}
    \frac{P \believes Q \said (X, Y)}{P \believes Q \said X}
    \label{eq:gny:i7} \tag{I7}
  \end{equation}

\end{mrule}

\begin{mrule}[Jurisdiction] Rules governing the meaning of jurisdiction.
  \begin{equation}
    \frac{P \believes Q \controls C, P \believes Q \believes C}{
      P \believes C
      } \label{eq:gny:j1} \tag{J1} 
  \end{equation}

  \begin{equation}
    \frac{P \believes Q \controls Q \believes *,
      P \believes Q \said (X \means C),
      P \believes \fresh{X} }{
      P \believes Q \believes C
      } \label{eq:gny:j2} \tag{J2} 
  \end{equation}

  \begin{equation}
    \frac{P \believes Q \controls Q \believes *,
      P \believes Q \believes Q \believes C}{
      P \believes Q \believes C
      } \label{eq:gny:j3} \tag{J3} 
  \end{equation}
\end{mrule}

\begin{definition}[Goals of Authentication] For an authentication there are 
  a number of possible goals:
  \begin{enumerate}
  \item Assurance: to assure another principal that a message has been
    received. $P \believes Q \sees X$.
  \item Exchanging secrets: to send to another principal a secret:
    \begin{equation*}
      \begin{aligned}
        P \believes P \secret{S} Q
      \end{aligned}
      \text{,} \quad
      \begin{aligned}
        Q \believes P \secret{S} Q
      \end{aligned}
    \end{equation*}
  \item Exchanging secrets: to send to another principal a secret and be
    assured of it:
    \begin{equation*}
      \begin{aligned}
        P \believes Q \believes P \secret{S} Q
      \end{aligned}
      \text{,} \quad
      \begin{aligned}
        Q \believes P \believes P \secret{S} Q
      \end{aligned}
    \end{equation*}
  \end{enumerate}
\end{definition}

\subsection{Example: The Kerberos Protocol}
\label{sec:kerb}

As an illustration of the use of the BAN logic, the
Kerberos\cite{sec:kerberos} protocol will be analyzed. This protocol is
supported within the emerging Transport Level Security specification. It is 
designed to establish a session key between two principals given a trusted
key server.

\begin{figure}[htbp]
  \begin{center}
  \includegraphics{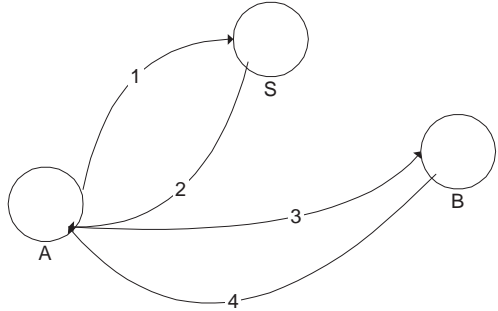}
  \caption{Kerberos Protocol: Messages}
  \label{fig:auth-1}
  \end{center}
\end{figure}

Referring to figure \ref{fig:auth-1}, the protocol uses three messages to
authenticate the sender and the fourth for mutual authentication of the
receiver to the sender. The messages appear in order below:

\begin{align}
  A \rightarrow S & \colon (A, B) \label{kerb:m1} \tag{$M_1$} \\
  S \rightarrow A & \colon
  \encrypted{T_S, L, K_{AB}, B,
    \encrypted{T_S, L, K_{AB}, A}{K_{BS}}}
  {K_{AS}}
  \label{kerb:m2} \tag{$M_2$} \\
  A \rightarrow B & \colon
  \encrypted{T_S,L,K_{AB},A}{K_{BS}},\encrypted{A,T_A}{K_{AB}}
  \label{kerb:m3} \tag{$M_3$} \\
  B \rightarrow A & \colon \encrypted{T_A + 1}{K_{AB}}
  \label{kerb:m4} \tag{$M_4$}
\end{align}

\begin{description}
\item[\ref{kerb:m1}] $A$ indicates that it wants a session key
with $B$ by sending two identifiers to $S$.
\item[\ref{kerb:m2}] $S$ returns a message that is encrypted under for
  $A$ only, it contains a timestamp, a session key, a restatement of
  the target for which it can be used and, the ingenious part, an
  encrypted message for $B$.
\item[\ref{kerb:m3}] $A$ sends the encrypted part on to $B$, which
  contains the timestamp, the session key and the other party to the
  session key. As well as that, $A$ sends a challenge, encrypted under 
  the session key.
\item[\ref{kerb:m4}] $B$ responds to the challenge by sending back the
  timestamp with a pre-determined calculation applied to it.
\end{description}

The timestamps $T_S$ and $T_A$ have a lifetime $L$. Effectively, these act
as nonces, so they shall be named as such: $N_S$ and $N_A$. There are some
preconditions about key distributions:

\begin{mySubequations}
  \begin{equation}
    \begin{aligned}
      A \believes A \secret{K_{AS}} S
    \end{aligned}
    \text{,} \quad
    \begin{aligned}
      B \believes B \secret{K_{BS}} S
    \end{aligned}
  \end{equation}

  \begin{equation}
    \begin{aligned}
      S \believes A \secret{K_{AS}} S
    \end{aligned}
    \text{,} \quad
    \begin{aligned}
      S \believes B \secret{K_{BS}} S
    \end{aligned}
  \end{equation}

  \begin{equation}
    S \believes A \secret{K_{AB}} B
  \end{equation}
\end{mySubequations}

They all hold long--term keys with each $S$ and vice--versa.

\begin{mySubequations}
  \begin{equation}
    \begin{aligned}
      A \holds K_{AS}
    \end{aligned}
    \text{,} \quad
    \begin{aligned}
      B \holds K_{BS}
    \end{aligned}
  \end{equation}
  \begin{equation}
    \begin{aligned}
      S \holds K_{AS}
    \end{aligned}
    \text{,} \quad
    \begin{aligned}
      S \holds K_{BS}
    \end{aligned}
  \end{equation}
\end{mySubequations}

There also some preconditions on jurisdiction and nonces:

\begin{mySubequations}
  \begin{equation}
    \begin{aligned}
      A \believes ( S \controls A \secret{K_{AB}} B )
    \end{aligned}
    \text{,} \quad
    \begin{aligned}
      B \believes ( S \controls A \secret{K_{AB}} B )
    \end{aligned}
  \end{equation}
  \begin{equation}
    \begin{aligned}
      A \believes \fresh{N_S}
    \end{aligned}
    \text{,} \quad
    \begin{aligned}
      B \believes \fresh{N_S}
    \end{aligned}
    \text{,} \quad
    \begin{aligned}
      B \believes \fresh{N_A}
    \end{aligned}
  \end{equation}
\end{mySubequations} 

Finally, the protocol can be analyzed: \eqref{kerb:m1} establishes no
new beliefs, but with \eqref{kerb:m2}, the following are established:
\begin{align*}
  M_2 \eqdef & \encrypted{(X, Y)}{K_{AS}} \\
  X \eqdef & N_S, (A \secret{K_{AB}} B) \\
  Y \eqdef & \encrypted{N_S, (A \secret{K_{AB}} B)}{K_{BS}} \\
  \Rightarrow & A \believes S \said (X, Y) \\
  \intertext{By initial key distribution, \eqref{eq:gny:t3} and
    \eqref{eq:gny:i1}}
  \droptext{From $X$}
  \Rightarrow & A \sees (N_S, (A \secret{K_{AB}} B)) \\
  \intertext {By \eqref{eq:gny:t3}}
  \Rightarrow & A \believes \fresh{A \secret{K_{AB}} B} \\
  \intertext{By \eqref{eq:gny:f1}}
  \Rightarrow & A \believes A \secret{K_{AB}} B, A \holds K_{AB} \\
  \intertext{By pre--conditions and \eqref{eq:gny:j1}}
\end{align*}

$B$ on receipt of \eqref{kerb:m3} can establish the following:
\begin{align*}
  M_3 & \eqdef (X, Y) \\
  \intertext{where}
  X & \eqdef \encrypted{N_S, (A \secret{K_{AB}} B)}{K_{BS}} \\
  Y & \eqdef \encrypted{N_A}{K_{AB}} \\
  \intertext{From $X$}
  \Rightarrow & B \believes S \said X \\
  \intertext{By initial key distribution and \eqref{eq:gny:i1}}
  \Rightarrow & B \sees (N_S, (A \secret{K_{AB}} B) \\
  \intertext {By \eqref{eq:gny:t3}}
  \Rightarrow & B \believes \fresh{A \secret{K_{AB}} B} \\
  \intertext{By \eqref{eq:gny:f1}}
  \Rightarrow & B \believes A \secret{K_{AB}} B, A \holds K_{AB} \\
  \intertext{By \eqref{eq:gny:t1} and jurisdiction pre--conditions}
  \droptext{From $Y$}
  \Rightarrow & B \believes A \said N_A \\
  \intertext{By \eqref{eq:gny:i1}}
  \Rightarrow & B \believes A \believes A \secret{K_{AB}} B,
  B \believes A \holds K_{AB} \\
  \intertext{By \eqref{eq:gny:i1}}
\end{align*}

$A$ on receipt of \eqref{kerb:m4} can establish the following:
\begin{align*}
  M_4 & \eqdef \encrypted{N_A}{K_{AB}} \\
  \intertext{where}
  \Rightarrow & A \sees{N_A} \\
  \intertext{By \eqref{eq:gny:i1}}
  \Rightarrow & A \believes B \believes A \secret{K_{AB}} B,
  A \believes B \holds K_{AB} \\
  \intertext {By \eqref{eq:gny:f1} and \eqref{eq:gny:i1}}
\end{align*}


\section{Funding and Author Details}

Research was funded by the Engineering and Physical Sciences Research
Council of the United Kingdom. Thanks to Malcolm Clarke, Russell--Wynn
Jones and Robert Thurlby.

\begin{verse}
  Walter Eaves \\
  Department of Electrical Engineering, \\
  Brunel University \\
  Uxbridge, \\
  Middlesex UB8 3PH, \\
  United Kingdom
\end{verse}

\begin{verse}
  \url{Walter.Eaves@bigfoot.com} \\
  \url{Walter.Eaves@brunel.ac.uk}
\end{verse}

\begin{verse}
  \url{http://www.bigfoot.com/~Walter.Eaves} \\
  \url{http://www.brunel.ac.uk/~eepgwde}
\end{verse}


\begin{thebibliography}{{CCI}88}

\bibitem[ABLP91]{sec:dist}
M~Abadi, M~Burrows, B~Lampson, and G~Plotkin.
\newblock A calculus for access control in distributed systems.
\newblock Technical report, DEC Systems Research Center, August 1991.
\newblock \myURL{ftp//ftp.digital.com}.

\bibitem[AG98]{sec:spi}
M~Abadi and Andrew~D Gordon.
\newblock A calculus for cryptographic protocols - the spi calculus.
\newblock Technical report, DEC Systems Research Center, January 1998.
\newblock \myURL{ftp//ftp.digital.com}.

\bibitem[BAN90]{BurrowsAbadiNeedham90}
Michael Burrows, Martin Abadi, and Roger Needham.
\newblock A logic of authentication.
\newblock {\em ACM Transactions on Computer Systems}, 8(1):18--36, February
  1990.


\bibitem[{CCI}88]{CCITTConsult88b}
{CCITT (Consultative Committee on International Telegraphy and Telephony)}.
\newblock {\em Recommendation ${X}.509$: The Directory---Authentication
  Framework}, 1988.

\bibitem[CER98]{sec:pkcs}
Pkcs.
\newblock World-Wide Web, July 1998.
\newblock \myURL{http://www.cert.org/advisories/CA-98.07.PKCS.html}.

\bibitem[DA97]{draft:tls}
Tim Dierks and Christopher Allen.
\newblock The {TLS} protocol: Version 1.0.
\newblock In Postel \cite{internet:draft}, November 1997.
\newblock
  \myURL{ftp://ietf.nri.reston.va.us/internet-drafts/draft-ietf-tls-protocol-0%
5.txt.Z}, Expires May 12, 1998.

\bibitem[DH77]{crypto:dh}
W.~Diffie and M.~E. Hellman.
\newblock New directions in cryptography.
\newblock {\em IEEE Transactions on Information Theory}, V IT-22(6), June 1977.


\bibitem[DS81]{Denning:1981:TKD}
D.~E. Denning and M.~S. Sacco.
\newblock Timestamps in key distribution protocols.
\newblock {\em Communications of the ACM}, 24(7):533--536, August 1981.


\bibitem[FKK95]{draft:ssl}
Alan~O. Freier, Philip Karlton, and Paul~C. Kocher.
\newblock The {SSL} protocol: Version 3.0.
\newblock In Postel \cite{internet:draft}, November 1995.
\newblock \myURL{http://www.netscape.com/eng/ssl3/draft302.txt}, Expired.

\bibitem[GNY90]{GoNeYa90}
Li~Gong, Roger Needham, and Raphael Yahalom.
\newblock Reasoning about belief in cryptographic protocols.
\newblock In {\em Symposium on Research in Security and Privacy}, pages
  234--248. {IEEE} Computer Society, {IEEE} Computer Society Press, May 1990.
\newblock \myURL{http://java.sun.com/people/gong/papers/gny-oakland.ps.gz}.

\bibitem[ITU89]{OSI:dir}
ITU.
\newblock The directory authentication framework.
\newblock In {\em Information technology - Open Systems Interconnection},
  number 509 in Series X Recommendations X.200 to X.900. International
  Telecommunication Union, International Telecommunication Union (ITU), Place
  des Nations, CH-1211 Geneva 20, Switzerland, 1989.
\newblock \myURL{http://info.itu.ch/itudoc/itu-t/rec/x/x200up.html}.

\bibitem[Low96]{Lowe96b}
Gavin Lowe.
\newblock Some new attacks upon security protocols.
\newblock In {\em Proceedings of the Computer Security Foundations Workshop
  VIII}. IEEE Computer Society Press, 1996.

\bibitem[NT94]{sec:kerberos}
B.~Clifford Neuman and Theodore Ts'o.
\newblock Kerberos: An authentication service for computer networks.
\newblock {\em IEEE Communications}, 32(9):33--38, September 1994.


\bibitem[Pos98]{internet:draft}
Jon Postel, editor.
\newblock {\em Internet Requests for Comment Drafts}.
\newblock Internet Executive Task-Force, \myURL{http://www.ietf.org}, August
  1998.


\bibitem[Ros95]{Rosche95}
R.~Martin Roscheisen.
\newblock General certificates.
\newblock World-Wide Web, August 1995.
\newblock
  \myURL{http://www-diglib.stanford.edu/cgi-bin/WP/get/SIDL-WP-1995-0012},
  \myURL{http://www-diglib.stanford.edu/diglib/WP/PUBLIC/DOC18.ps}, v 0.9.

\bibitem[THA99]{ca:thawte}
Thawte digital certificate services.
\newblock World--Wide Web, Mar 1999.
\newblock \myURL{http://www.thawte.com}.

\bibitem[XZX97]{sec:xu}
Shonhuai Xu, Gendu Zhang, and Hong Xhu.
\newblock On the properties of cryptographic protocols the weaknesses of
  ban-like logics.
\newblock {\em ACM Operating Systems Review}, 31(4):12--24, October 1997.


\end{thebibliography}
\end{document}